\documentclass[aip,jap,reprint,showpacs,superscriptaddress]{revtex4-1}

\usepackage{hyperref}
\usepackage{graphicx}
\usepackage{dcolumn}

\begin{document}

\noindent {\textbf{The following article appeared in the Journal of Applied Physics and may be found at \href{http://link.aip.org/link/?jap/113/073906}{P.J. Metaxas et al., J. Appl. Phys. 113, 073906 (2013)}. Copyright 2013 American Institute of Physics. This article may be downloaded for personal use only. Any other use requires prior permission of the author and the American Institute of Physics.}}

\vspace{1cm}

\title{Spatially periodic domain wall pinning potentials: Asymmetric pinning and dipolar biasing}

\author{P.J.~Metaxas}
\email{peter.metaxas@uwa.edu.au}
\affiliation{School of Physics, M013, University of Western Australia, 35 Stirling Hwy, Crawley WA 6009, Australia.}
\affiliation{Laboratoire de Physique des Solides, Universit\'e Paris-Sud 11, CNRS, UMR 8502, F-91405 Orsay Cedex, France.
}%
\affiliation{
Unit\'e Mixte de Physique CNRS/Thales, 1 Avenue Augustin Fresnel, 91767 Palaiseau, France and Universit\'e Paris-Sud 11, 91405 Orsay, France.}

\author{P.-J.~Zermatten}
\affiliation{
SPINTEC, UMR-8191, CEA-INAC/CNRS/UJF-Grenoble 1/Grenoble-INP, 17 rue des Martyrs, 38054 Grenoble
cedex 9, France.
}

\author{R.L.~Novak}%
\author{S.~Rohart}%
\author{J.-P.~Jamet}%
\author{R.~Weil}%
\author{J.~Ferr\'{e}}%
\author{A.~Mougin}%
\affiliation{Laboratoire de Physique des Solides, Universit\'e Paris-Sud 11, CNRS, UMR 8502, F-91405 Orsay Cedex, France.
}%

\author{R.~L. Stamps}
\affiliation{School of Physics, M013, University of Western Australia, 35 Stirling Hwy, Crawley WA 6009, Australia.}
\affiliation{
SUPA - School of Physics and Astronomy, University of Glasgow, G12 8QQ Glasgow, United Kingdom.
}%

\author{G.~Gaudin}
\author{V.~Baltz}
\author{B.~Rodmacq}
\affiliation{
SPINTEC, UMR-8191, CEA-INAC/CNRS/UJF-Grenoble 1/Grenoble-INP, 17 rue des Martyrs, 38054 Grenoble
cedex 9, France.
}%

\date{\today}

\begin{abstract}
Domain wall propagation has been measured in continuous, weakly disordered, quasi-two-dimensional, Ising-like magnetic layers that are subject to spatially periodic domain wall pinning potentials. The potentials are generated non-destructively using the stray magnetic field of ordered arrays of magnetically hard [Co/Pt]$_m$ nanoplatelets which are patterned above and are physically separated from the continuous magnetic layer. The effect of the periodic pinning potentials on  thermally activated domain wall creep dynamics is shown to be equivalent, at first approximation, to that of a uniform, effective retardation field, $H_{ret}$, which acts against the applied field, $H$. We show that $H_{ret}$ depends not only on the array geometry but also on the relative orientation of $H$ and the magnetization  of the nanoplatelets. A result of the latter dependence is that wall-mediated hysteresis loops obtained for a set nanoplatelet magnetization  exhibit many properties that are normally associated with ferromagnet/antiferromagnet exchange bias systems. These include a switchable bias, coercivity enhancement and  domain wall roughness that is dependent on the applied field polarity. 
\end{abstract}


\newcommand{\hup}{$H\uparrow$}
\newcommand{\hdown}{$H\downarrow$}
\newcommand{\arrayup}{$M_{array}\uparrow$}
\newcommand{\arraydown}{$M_{array}\downarrow$}
\newcommand{\arrayzero}{$M_{array}=0$}
\newcommand{\up}{$\uparrow$}
\newcommand{\down}{$\downarrow$}
\newcommand{\marray}{$M_{array}$}
\newcommand{\hret}{$H_{ret}$}
\newcommand{\hretp}{$H^{P}_{ret}$}
\newcommand{\hretap}{$H^{AP}_{ret}$}
\newcommand{\tstep}{$t_{step}$}

\maketitle

\section{Introduction}

Controllable pinning of magnetic domain walls will be a key enabler for domain wall-based spintronic  devices such as  memristors\cite{Wang2009,Munchenberger2012}, shift registers\cite{Allwood2002,Allwood2003,Allwood2006,Hayashi2008science,Kim2010,Franken2012} and binary memory cells\cite{Fukami2008,Fukami2009,Honjo2012}. In such devices domain walls are moved through what are generally sub-micron ferromagnetic strips wherein the state of the device is related to the physical position of the domain wall. Stable domain wall positioning is often achieved via strong local structural modifications to the strip such as the introduction of notches\cite{McMichael2000,Himeno2003,Klaui20032,Briones2008,hayashi2008} or holes\cite{RodriguezRodriguez2007,PerezJunquera2008,Alija2011,Marconi2011}. Such techniques have the advantage of allowing for the definition of the strip and the domain wall pinning sites in a single lithographic step. However, there can be challenges associated with the fabrication of multiple notches with identical pinning properties as well as non-trivial pinning effects at notches in multilayer devices\cite{Briones2008}. 

An alternative pinning method involves the definition of well-localized regions within the strip where the local (effective) magnetic field differs to that in the surrounding regions. This can be done via local modification of the strip's anisotropy \cite{Franken2012} or by using exchange\cite{FraileRodriguez2006} or dipolar\cite{Metaxas2009,OBrien2011} coupling between the strip and  suitably placed sub-micron magnetic elements. Here we build on previous work\cite{Metaxas2009} where we studied spatially periodic pinning potentials generated by dipolar coupling between a single periodic array of 0.2 $\mu$m wide, magnetically hard nanoplatelets (period = 1.2 $\mu$m) and a softer underlying continuous ultrathin ferromagnetic Co layer. In this  work we study arrays with inter-platelet spacings as small as $\sim$0.1 $\mu$m. The reduced inter-platelet spacings result in stronger nanoplatelet-induced reductions of the wall velocity within the thermally activated low field wall motion regime as well as increased  pinning asymmetry  (see below).  Although we focus on pinning effects in a continuous magnetic film, the lithographic methods can be extended to strips.

In both our previous and current studies, a 5 nm Pt spacer separates the nanoplatelets from the underlying continuous film  which ensures that coupling between the nanoplatelets and the film is  primarily dipolar in nature.  This allows for the periodic pinning potential to be introduced without significant modification to the physical structure of the   ultrathin Pt/Co-based film below which ensures that the film's intrinsic weak disorder is not modified. This means that it is possible to study the effects of two independent but coexistent domain wall pinning potentials, one periodic, the other disordered. Since domain walls in Pt/Co/Pt films are described well by general theories for elastic interface motion in disordered systems\cite{Lemerle1998,Metaxas2007,Kim2009}, there is a fundamental interest in creating model experimental systems for studying interface motion where pinning effects can be controllably modified.

One of the primary focuses of this article however will be pinning asymmetry: the nanoplatelet-induced pinning effects depend upon the relative orientation of the nanoplatelet magnetizations and the applied field, $H$, that is used to drive domain wall propagation in the continuous layer. 
Beyond providing field-polarity-dependent domain wall propagation (and therefore tunable pinning), the asymmetry  has a further consequence: for a given orientation of the nanoplatelet magnetizations,  domain wall mediated switching of the  continuous layer's magnetization results in a magnetic hysteresis loop which is not centered around $H=0$, ie. it exhibits a `dipolar bias'. Indeed, characteristics of this nanoplatelet-film system will be shown to be highly reminiscent of those seen in conventional ferromagnet/antiferromagnet exchange bias systems\cite{Meiklejohn1956,Nogues1999,Stamps2000,Kiwi2001}.

\section{Film-nanoplatelet structures and pinning}\label{samples}

\subsection{Fabrication}

To fabricate  arrays  of nanoplatelets above a continuous film, continuous multilayer Co/Pt stacks were first deposited by dc sputtering on Si/SiO$_2$ substrates which were initially cleaned using an Argon plasma. Starting from the substrate, each film consists of a Pt buffer layer, a Co layer or Co/Pt/Co stack, a Pt spacer layer and an upper Co/Pt multilayer stack (see Table 1). The individual Co layers within each  stack are coupled ferromagnetically across thin Pt layers (1.6 or 1.8 nm) which  ensures a cooperative magnetic reversal of the layers  \cite{Baltz2005,Wiebel2005,Metaxas2009,SanEmeterioAlvarez2010}. The lower, ultrathin, magnetically soft Co layer structure is separated from the upper stack by a 5nm thick Pt spacer layer. As such, the effective fields\cite{Moritz2004} associated with interlayer exchange and dipolar `orange peel' coupling  are weak (on the order of or less than 1 Oe) and in the final structure are dominated by the strong dipolar fields of the nanoplatelets (see Sec.~\ref{sec_pinning}). 

\begin{figure}[htbp]
\includegraphics[width=7cm]{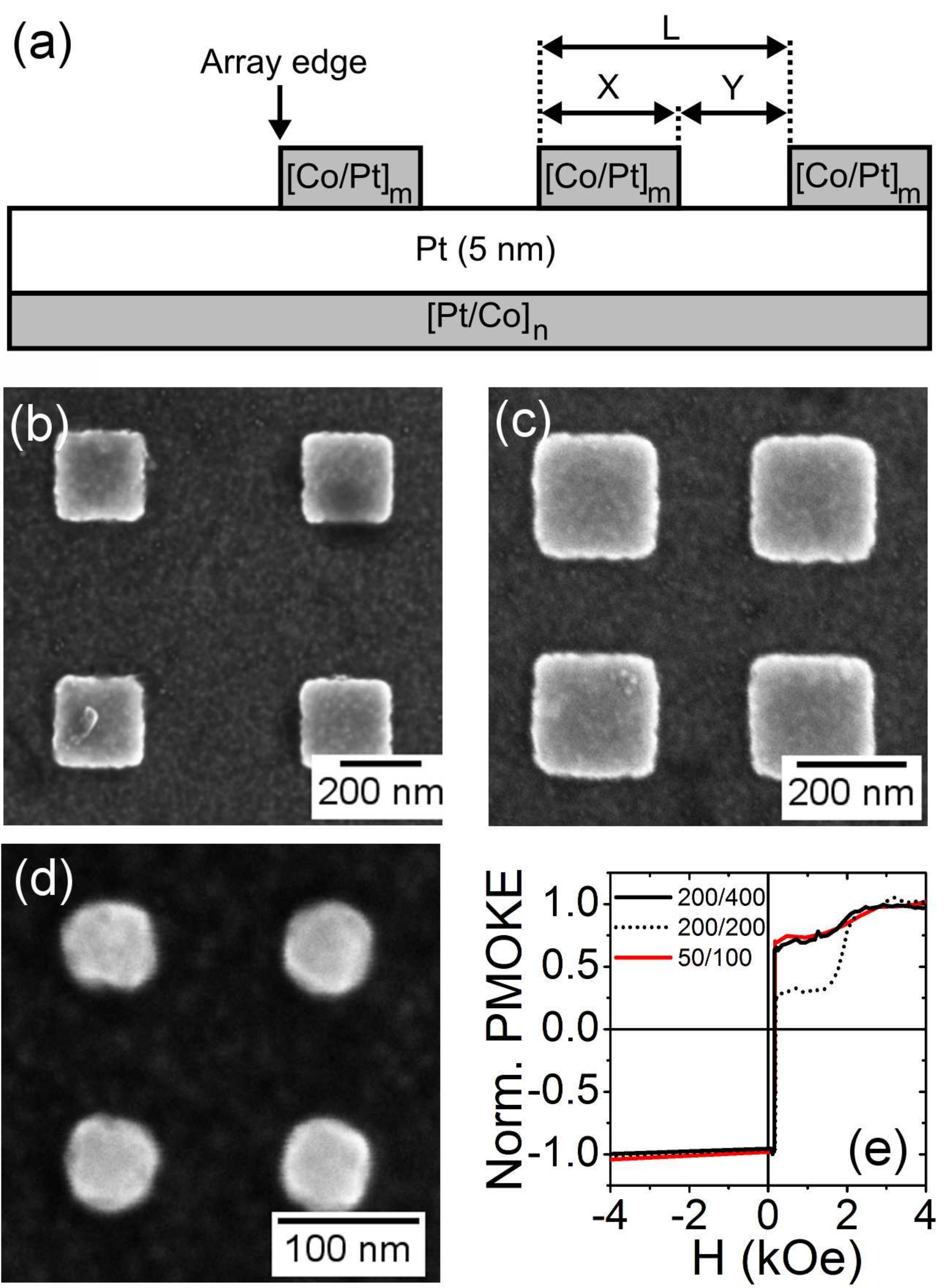}
\caption{(Color online) (a) Cross-sectional schematic showing the general structure of each sample. The lower continuous soft ferromagnetic film and upper hard multilayer nanoplatelet  stacks  are shaded in gray. Each array is denoted X/Y with X and Y respectively the intended width and edge-to-edge spacing of the nanoplatelets. Zoomed in scanning electron micrographs of the (b) 200/400 and (c) 200/200 and (d) 50/100 nanoplatelet arrays. (e) Positive-field-driven switching of the continuous film (sharp transition) and nanoplatelets (gradual transition) following negative saturation measured using the spatially averaged polar magneto-optical Kerr effect (PMOKE) microscopy method outlined in Sec.~\ref{sec_hyst}.}
\label{fig_1}
\end{figure}

The magnetically hard, upper Co/Pt multilayer stack was patterned to define 50 and 100 $\mu$m wide arrays of nanoplatelets  [Fig.~\ref{fig_1}(a)]. To do this, a Titanium mask was first fabricated on top  of the film using electron beam lithography and then the entire sample was etched with a low energy (150 eV) unfocussed Argon ion beam. 
Etching of the upper stack was monitored using secondary ion mass spectrometry (SIMS) and was halted upon reaching the Pt spacer. In this way, arrays of hard ferromagnetic nanoplatelets could be defined whilst preserving the structure of the lower continuous layer. The upper stack was completely removed in the regions surrounding each array.  Four array geometries have been studied which had intended nanoplatelet size/spacing values (in nm) of X/Y =  200/1000 (see Ref.~\onlinecite{Metaxas2009}), 200/200, 200/400, 50/100. The actual values of X and Y,  measured using scanning electron microscopy [eg.~Figs.~\ref{fig_1}(b-d)], are given in Table 1 together with the film and nanoplatelet compositions.

\begin{table*}[htbp]
{\scriptsize
\begin{tabular}{|c|c|c|c|c|c|}
\hline 
Intended 	&Actual nanoplatelet &Actual nanoplatelet & Continuous layer  & Spacer  & Nanoplatelet\\
X/Y value 	&width (X) & separation (Y) 				&   composition 				& layer & composition\\
\hline
200/1000	& 230 & 940 & [Pt(1.8)/Co(0.6)]$_2$  & Pt(5) & [Co(0.6)/Pt(1.8)]$_4$ \\
\hline
200/400	& 225 & 377 & Pt/Co(0.65)  & Pt(5) & Co(0.7)/[Pt(1.6)/Co(0.55)]$_4$/Pt \\
\hline
200/200	& 234 & 172 & Pt/Co(0.65)  & Pt(5) & Co(0.7)/[Pt(1.6)/Co(0.55)]$_4$/Pt \\
\hline
50/100	& 62 & 89 & Pt/Co(0.65)  & Pt(5) & Co(0.7)/[Pt(1.6)/Co(0.55)]$_4$/Pt \\
\hline 
\end{tabular} }
\caption{ Nanoplatelet array characteristics (see Fig.~\ref{fig_1}(a)). All lengths and film thicknesses are in nm. The actual nanoplatelet widths, X, and edge-to-edge separations, Y,  were measured by scanning electron microscopy [eg.~Figs.~\ref{fig_1}(b-d)].} 
\label{tab_exp}
\end{table*}

Switching of the 200/400, 200/200 and 50/100 nanoplatelet arrays and continuous film were measured magneto-optically [Fig.~\ref{fig_1}(e), see caption for measurement details]. Both the continuous film and  nanoplatelets retain a perpendicular magnetic anisotropy after patterning with the nanoplatelets displaying a rather broad distribution\cite{Jamet1998} of switching fields lying approximately between  1.5 kOe and 2.5 kOe. These relatively high switching fields are  advantageous here since they	prevent unintended reversal of the nanoplatelets  due to out of plane stray fields generated by neighboring nanoplatelets ($<100$ Oe, see Sec.~\ref{sec_pinning}) or the fields used to drive domain wall motion ($\sim$1 kOe at most).

\subsection{Domain wall pinning}\label{sec_pinning}

A domain wall moving in a region of the continuous layer that is located beneath an array will be subject to two sources of pinning. The first source of pinning arises from an interaction between the domain wall and the random, weak disorder that is intrinsic to continuous Pt/Co layers\cite{Lemerle1998,Metaxas2007}. This leads to the thermally activated creep dynamics that are typically observed in such films\cite{Lemerle1998,KrusinElbaum2001,Repain2004,Metaxas2007,SanEmeterioAlvarez2010}. 
The second source of pinning  arises from the dipolar field generated by the nanoplatelets which can locally impede domain wall motion\cite{Metaxas2009}. In the following, we discuss how the nanoplatelets' stray dipolar fields can locally aid or counteract the effect of the applied field which is always perpendicular to the film plane.

\begin{figure*}[htbp]
\includegraphics[width=17cm,clip=yes]{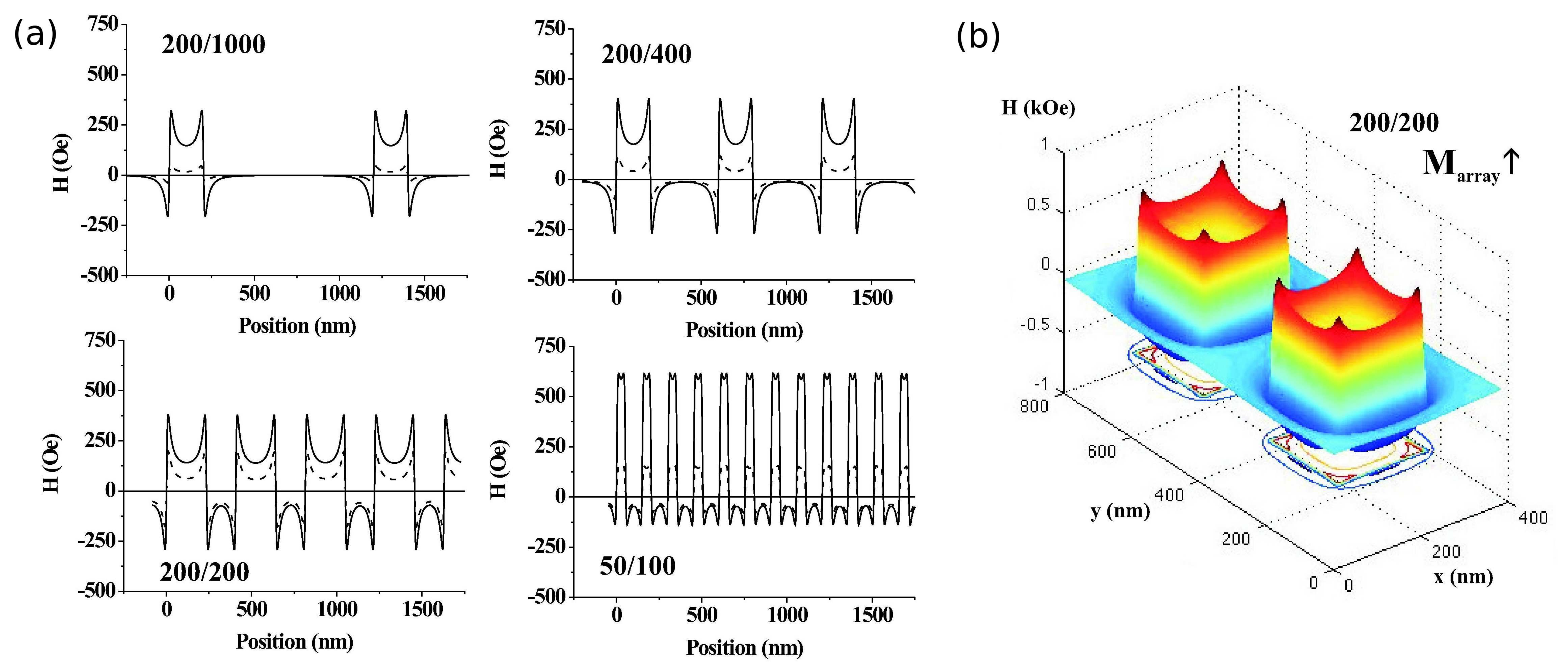}
\caption{(Color online)  (a) Perpendicular component of the stray field beneath positively saturated  nanoplatelets ($\uparrow$) as calculated at the center of the soft layer in the 200/1000, 200/400, 200/200 and 50/100 arrays. The solid line represents the stray field value along a line passing through the center of a row of nanoplatelets. The dotted line corresponds to its value averaged over a moving window in the plane of the continuous layer with a width of 10 nm and a length equal to the array period. The latter  allows one to visualize the field experienced by a straight wall moving through the array averaged along its length. (b) Plot of the perpendicular component of the nanoplatelet stray field for the 200/200 array beneath two neighboring nanoplatelets.}
\label{fig_field}
\end{figure*}

In Fig.~\ref{fig_field} we have plotted the perpendicular component of the stray  field, $H^Z_{stray}$, of positively magnetized nanoplatelets at the center of the soft continuous layer   as calculated along a line passing through the center of a row of nanoplatelets [Fig.~\ref{fig_field}(a)]  and  in two dimensions for the 200/200 array [Fig.~\ref{fig_field}(b)]. Directly beneath each nanoplatelet, $H^Z_{stray}$  is strong and  aligned with the nanoplatelet's magnetization direction. The highest spatial uniformity of this part of the stray field profile is achieved for small dot sizes. In the regions surrounding each nanoplatelet, $H^Z_{stray}$, is weaker and aligned antiparallel to the nanoplatelet magnetization. It changes sign just below the nanoplatelet border and exhibits a large gradient over a distance which is comparable to the domain wall width ($\sim 10$ nm)\cite{Metaxas2007}. At this location, the in-plane component of the stray field becomes large.

\begin{figure}[h]
\includegraphics[width=7cm,clip=yes]{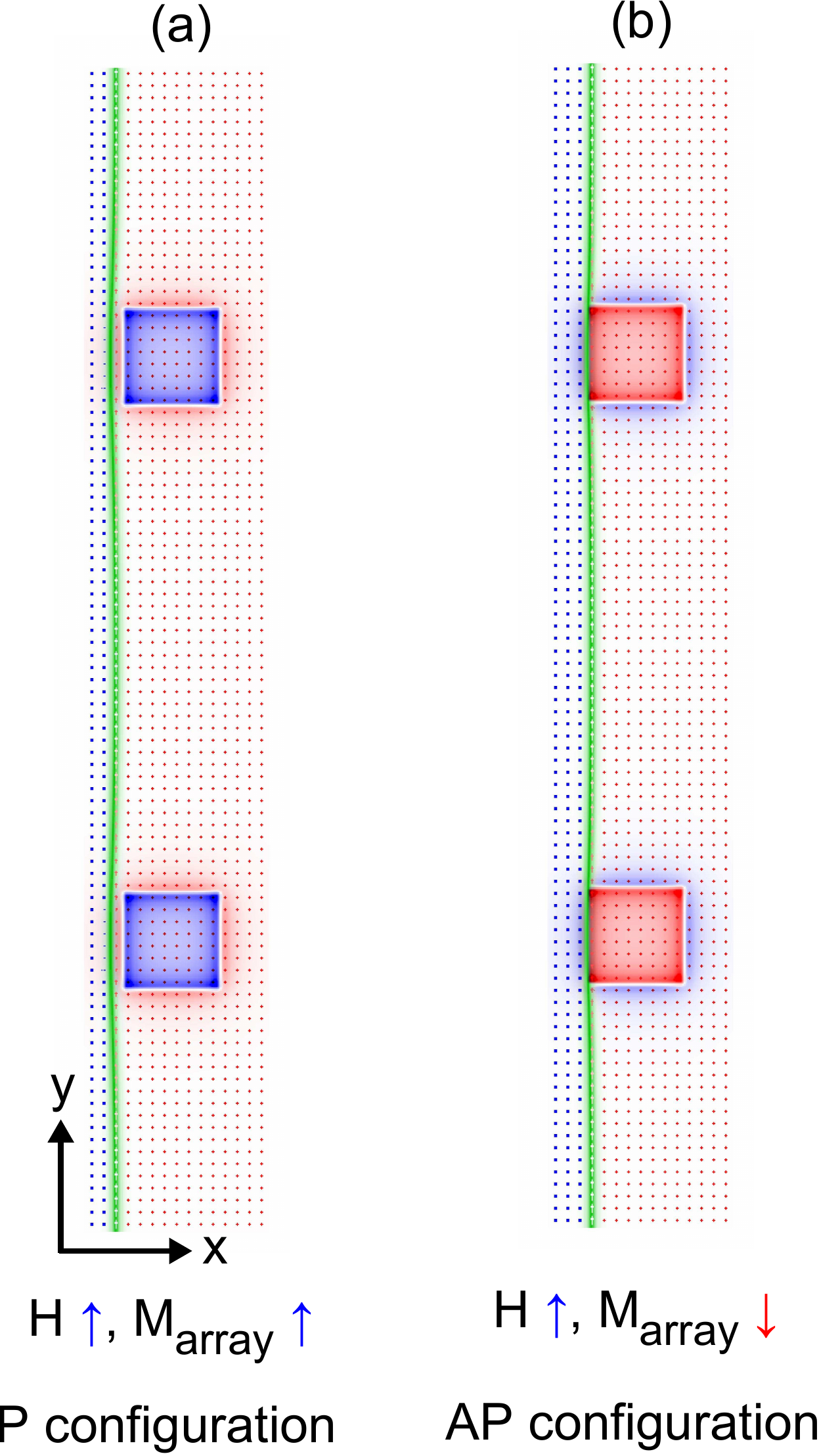}
\caption{(Color online) Micromagnetic simulation (400 nm $\times$  2400 nm) of domain walls propagating through an ideal (defect free) Co(0.65nm) layer towards the location of two nanoplatelets for the (a) P configuration ($M_{array}\uparrow$ and $H\uparrow$) and the (b) AP configuration ($M_{array}\downarrow$ and $H\uparrow$). The perpendicular component of the field of the nanoplatelets within the calculation region has been superimposed on the image (the blue and red `squares'). The color code for the film's magnetization and the nanoplatelets' perpendicular dipolar fields is +z=blue, -z=red and +y=green (Bloch domain wall).}
\label{fig_oommf}
\end{figure}

Depending on the  position of a domain wall relative to a nanoplatelet, $H^Z_{stray}$ can thus either  reinforce $H$ or oppose $H$ with the latter resulting in localized domain wall pinning.  For a given magnetization state of the nanoplatelets in the array, \marray~(\up~or \down), the positions which impede wall motion (and thus the pinning potential itself) will depend upon the sign of $H$. A consequence of this asymmetry can be directly visualized in  Fig.~\ref{fig_oommf} where we show simulated configurations of domain walls driven under positive field towards  positively or negatively magnetized nanoplatelets (see Appendix \ref{micromag-appendix} for simulation details).  For positively magnetized nanoplatelets [Fig.~\ref{fig_oommf}(a)], the wall stops  at a finite distance from the nanoplatelet border, repelled by the negative $H^Z_{stray}$ which surrounds the nanoplatelet. In contrast, for negatively magnetized nanoplatelets [Fig.~\ref{fig_oommf}(b)], the positive $H^Z_{stray}$ around the nanoplatelet  attracts the domain wall to the nanoplatelet's edge. In the following section we examine the consequences of these distinct pinning potentials on domain wall propagation and magnetic reversal.

\section{Magnetization reversal: experimental results}\label{sec_exp}

\subsection{Magnetic domain structures}\label{sec_moke}

\begin{figure}[htbp]
\includegraphics[width=5cm,clip=yes]{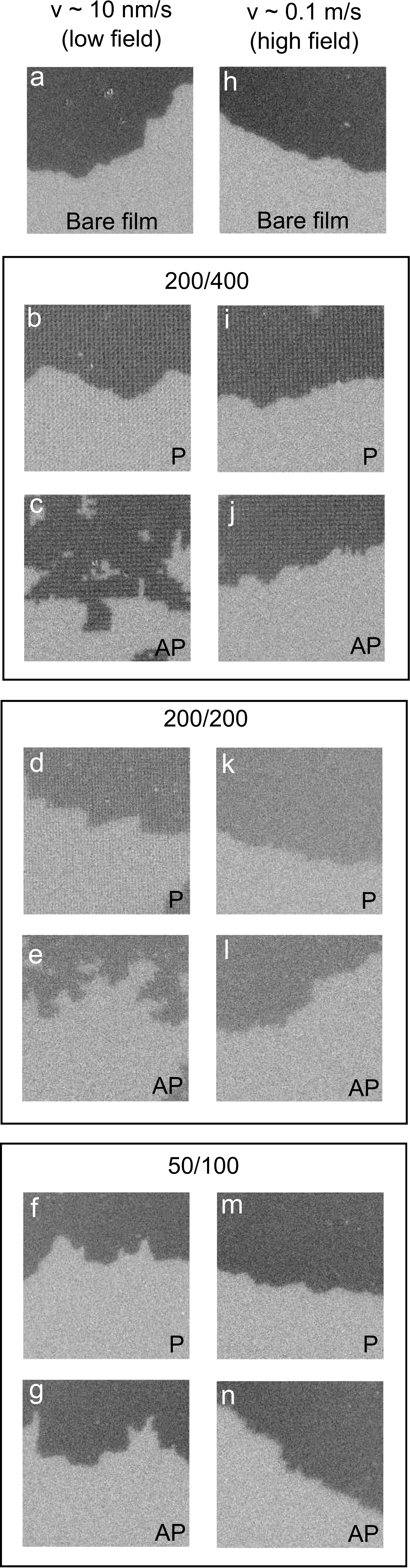}
\caption{Stabilized domains in the continuous ferromagnetic layer outside (a,h) and beneath the 200/400 (b,c,i,j), 200/200 (d,e,k,l) and 50/100 (f,g,m,n) arrays following wall propagation under fields parallel (P) or antiparallel (AP) to the magnetization of the nanoplatelets. The images in the two columns correspond to approximate wall velocities of 10 nm/s (a-g) and 0.1 m/s (h-n). The dark domains are propagating  towards the bottom of each image which has dimensions of 21.6 $\mu$m $\times$ 21.6 $\mu$m. }
\label{fig_moke}
\end{figure}

Domain structures in the continuous layer (both below and outside the arrays) were imaged quasi-statically using magnetic force microscopy (MFM) and a high resolution ($\sim$0.4  $\mu$m) PMOKE microscope. The two methods are complementary: PMOKE microscopy is rapid and gives reasonable resolution over tens of microns whereas  MFM, despite slower image acquisition, allows for an accurate determination of domain wall positioning relative to single nanoplatelets. Two primary magnetic configurations were investigated, parallel (P) and antiparallel (AP), referring to the relative orientation of the magnetization direction of the nanoplatelets in the array and the field used to drive domain wall motion beneath the array. We will also discuss some results obtained for a `demagnetized' array (D), where there are approximately equal numbers of $\uparrow$ and $\downarrow$ magnetized nanoplatelets.

PMOKE images of domain walls outside and beneath the P- and AP-configured 200/400, 200/200 and 50/100 arrays  can be seen in Fig.~\ref{fig_moke}.  The images were obtained in zero field following field-driven propagation at velocities on the order of $10^{-8}$ m/s [Figs.~\ref{fig_moke}(a-g)] and $10^{-1}$ m/s [Figs.~\ref{fig_moke}(h-n)] (see the following section for quantitative information on velocity-field characteristics).  Equivalent images for the 200/1000 array can be found in Ref.~\onlinecite{Metaxas2009}. Appendix \ref{moke-appendix} details the experimental protocols for preparing the system's magnetic configuration and driving domain wall motion. 

At low field [$v\sim 10^{-8}$ m/s; Figs.~\ref{fig_moke}(a-g)], except for the 200/400 array in the P configuration [Fig.~\ref{fig_moke}(b)], walls within the arrays are visibly different to those outside the arrays and there are observable differences between domain wall roughness beneath P- and AP-configured arrays.  For the 200/Y arrays in the AP configuration [Figs.~\ref{fig_moke}(c,e) and Ref.~\onlinecite{Metaxas2009}], small unreversed domains remain behind the expanding domain wall fronts which are  rougher than both those outside the arrays and those beneath the P-configured 200/Y arrays [Figs.~\ref{fig_moke}(b,d)]. These strong modifications can be explained by the fact that pinning in the AP configuration comes from strong stray fields existing directly beneath each nanoplatelet.  There are few clear differences in roughness for P and AP configurations in the 50/100 array [Figs.~\ref{fig_moke}(f,g,m,o)] however this may be due to limitations of the microscope's resolution.  at higher fields ($v\sim 10^{-1}$ m/s), walls beneath and outside the P- and AP-configured arrays exhibit only minimal differences  [Figs.~4(h-o)]. No clear dependence of domain wall roughness on the field polarity  in the demagnetized (D) configuration was identified (see Sec.~\ref{sec_hyst}).

In  Fig.~\ref{fig_mfm}, we show MFM images (see Appendix \ref{mfm-appendix} for experimental parameters) obtained for domain walls beneath the 200/1000 array (the more closely spaced nanoplatelets in the other arrays obscured the underlying domain structure and prevented MFM imaging). Many qualitative similarities with the PMOKE images can be noted. There is roughness that can be correlated with nanoplatelet positions and, in the AP configuration, the  unreversed domains left behind the propagating domain front are clearly pinned at their borders by the nanoplatelets. We also find positions where the nanoplatelet-domain wall positioning clearly mimic the pinned configurations predicted by the simulations in Sec.~\ref{sec_pinning}. Insets in both Figs.~\ref{fig_mfm}(a) and (b) show walls which stop at a finite distance from the nanoplatelet (P configuration, Fig.~\ref{fig_mfm}(a)/Fig.~\ref{fig_oommf}(a)) and those which hug the nanoplatelet borders (AP configuration, Fig.~\ref{fig_mfm}(b)/Fig.~\ref{fig_oommf}(b)).  Finally, the contrast levels of the nanoplatelets in Fig.~\ref{fig_mfm} demonstrate the uniformity of the remanent nanoplatelet magnetization states in the P and AP configurations [Fig.~\ref{fig_mfm}(a,b)] and their randomness in the D state [Fig.~\ref{fig_mfm}(c)].

\begin{figure*}[htbp]
\includegraphics[width=16cm,clip=yes]{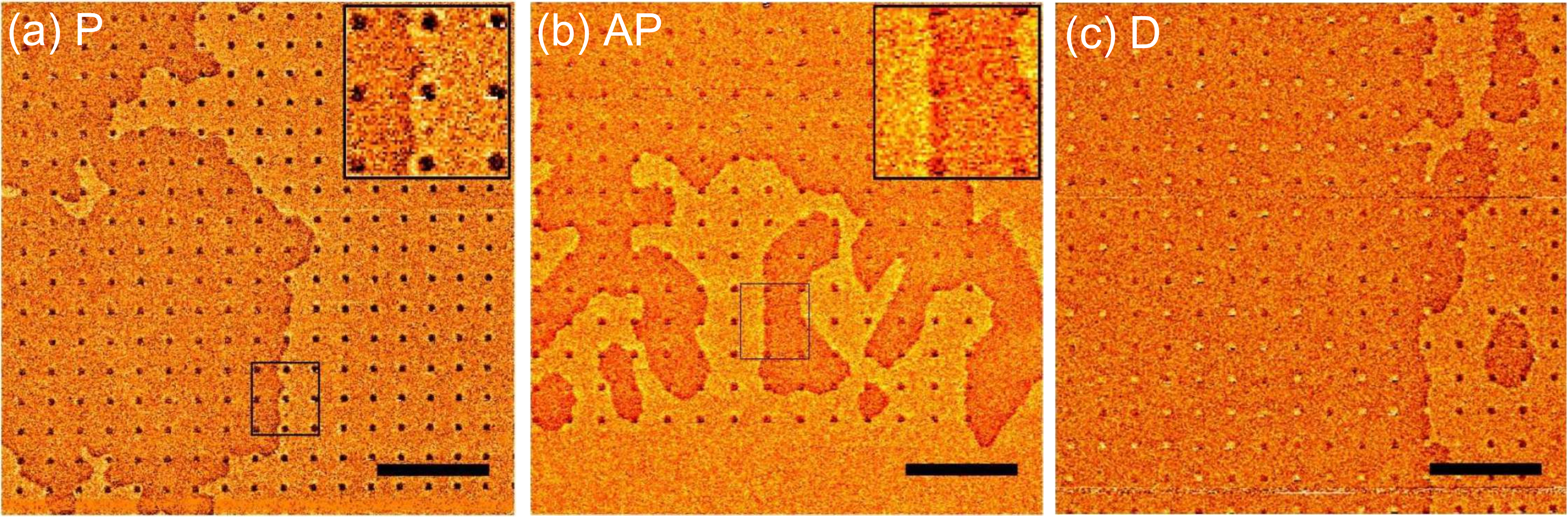}
\caption{(Color online)  MFM images of the 200/1000 array for driving fields (a) parallel, P ($H\uparrow$), or (b) antiparallel, AP ($H\downarrow$), to the nanoplatelet magnetization ($M_{array}\uparrow$), and in the (c) demagnetized D state ($M_{array} = 0$, \hup) where there are both \up~and \down~ magnetized nanoplatelets. The magnetic field used to drive the walls through the array was +130 Oe, -100 Oe and +80 Oe, for the P, AP and D configurations respectively. For the P state, the dark domain is expanding at the expense of the light domain. The inverse applies for the AP and D states. The length of each black scale bar is 4 $\mu$m. Magnified images of the regions inside the black squares are shown in the insets.}
\label{fig_mfm}
\end{figure*}

\subsection{Domain wall velocity in the creep regime}\label{sec_dynamics}

\begin{figure*}[htbp]
\includegraphics[width=16cm,clip=yes]{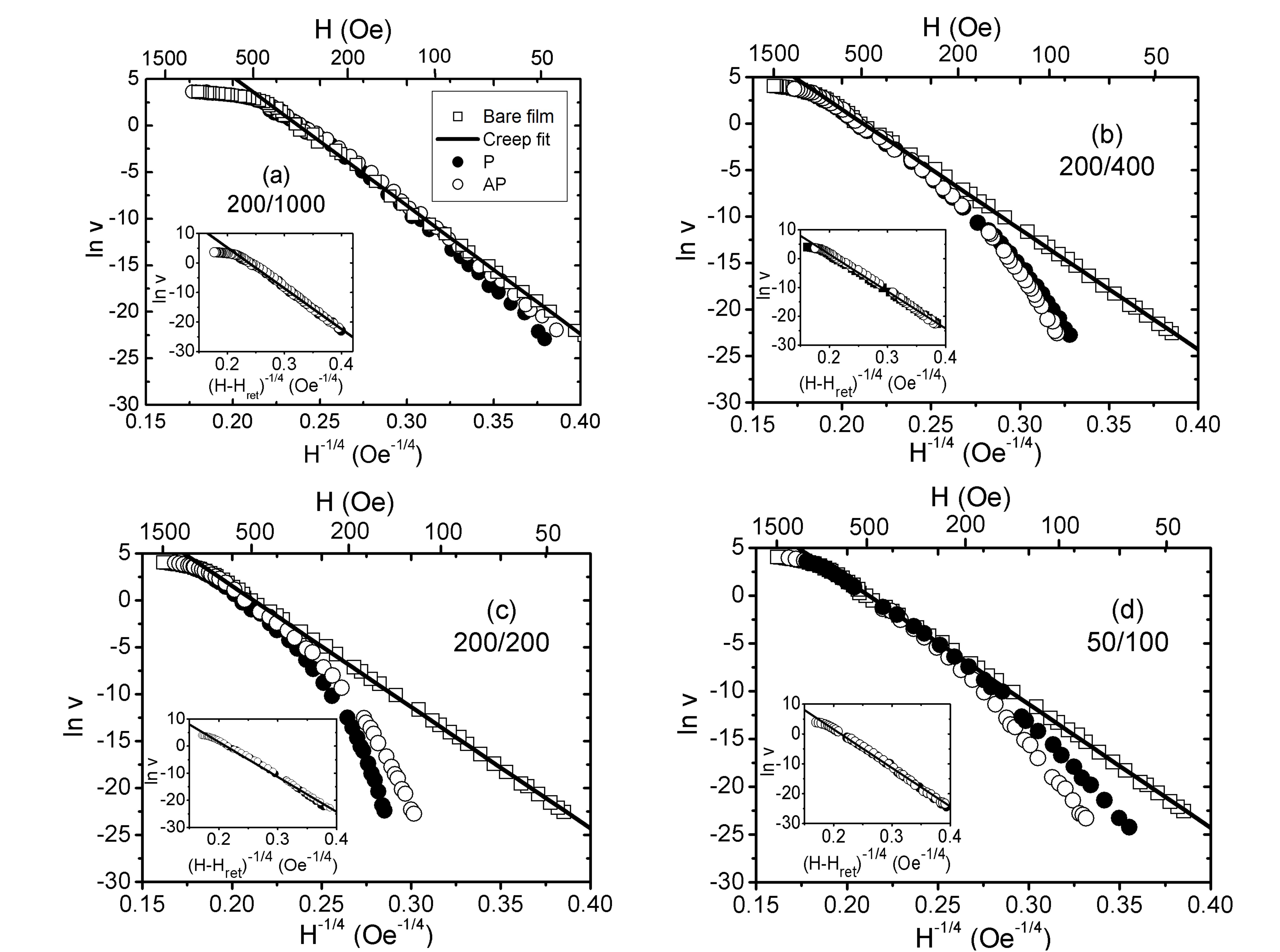}
\caption{Natural logarithm of the velocity, $v$, of domain walls in the continuous soft layer outside the arrays (`bare film') and beneath the arrays for P, AP and D configurations plotted against $H^{-1/4}$ for the (a) 200/1000 (data from Ref.~\onlinecite{Metaxas2009}), (b) 200/400, (c) 200/200, and (d) 50/100 arrays. Inset: the P and AP velocity curves can be superimposed on those measured in the absence of the nanoplatelets by plotting $\ln v$ versus $(H-H_{ret})^{-1/4}$. Data in (a) is  reprinted with permission from Appl.~Phys.~Lett.~\textbf{98}, 132504 (2009). Copyright 2009 American Institute of Physics. }
\label{fig_vels}
\end{figure*}

Field dependent domain wall velocities beneath and outside the arrays were determined using an established quasi-static  PMOKE microscopy technique\cite{Lemerle1998,Jamet1998,Metaxas2007} (Appendix \ref{moke-appendix}). Consistent with previous studies on ultrathin Pt/Co/Pt films and multilayers, the measured mean domain wall velocity, $v$, in the absence of the nanoplatelet arrays is well fitted by a creep law 
\begin{equation} 
v = v_0  \exp \left[- \frac{T_{dep}}{T} \left(\frac{H_{dep}}{H}\right)^{\mu}\right],\label{eq_creep}
\end{equation} 
consistent with the film's weak disorder and their quasi-two-dimensional, Ising-like nature.
$T_{dep}$ and $H_{dep}$ characterize the pinning potential associated with the film's structural disorder and $\mu=\frac{1}{4}$ is the dynamic universal exponent for a 1D interface moving through a 2D weakly disordered medium. The velocity of walls moving outside the arrays (marked as `bare film') is plotted in Fig.~\ref{fig_vels}  as $\ln v$ versus $H^{-1/4}$ with linearity at small $H$ (large $H^{-1/4}$) confirming robust agreement with Eq.~(\ref{eq_creep}) over ten orders of magnitude of $v$.

In the creep regime, the domain wall velocities beneath the arrays are consistently lower than those measured outside the arrays for both P and AP configurations. This pinning induced velocity reduction is manifested in a clear curvature or non-linearity in the $\ln v$ versus $H^{-1/4}$ data which signifies a departure  from the standard creep law which can be understood to a first approximation using a modified version\cite{Metaxas2009} of Eq.~(\ref{eq_creep}). Indeed, despite the complex spatial profile of the periodic pinning potentials, the wall velocity beneath each studied array can be fitted well by replacing $H$ in Eq.~(\ref{eq_creep}) with $H - H_{ret}$ where $H_{ret}>0$ (insets of Fig.~\ref{fig_vels}):
\begin{equation} 
v = v_0  \exp \left[- \frac{T_{dep}}{T} \left(\frac{H_{dep}}{H-H_{ret}}\right)^{\mu}\right].\label{eq_creep2}
\end{equation} 
The creep parameters intrinsic to the weakly disordered layer are left unchanged.
This approach linearizes the low field velocity data obtained beneath the arrays and provides good overlap with the creep data obtained outside the arrays. While the adjusted data is not perfectly linear [a very slight curvature can be seen under close inspection of the insets of Figs.~\ref{fig_vels}(b-d)], this is a reasonable first order model. The attractive aspect of Eq.~(\ref{eq_creep2}) is that it can be linked with a simple physical picture: the reduction in the measured creep velocity  arises from an effective retardation field, \hret, which always opposes the applied field, $H$. As a consequence, walls will be frozen for fields below $H_{ret}$ due to the presence of the nanoplatelets' stray fields. 
 
\begin{figure}[h]
\includegraphics[width=7.5cm,clip=yes]{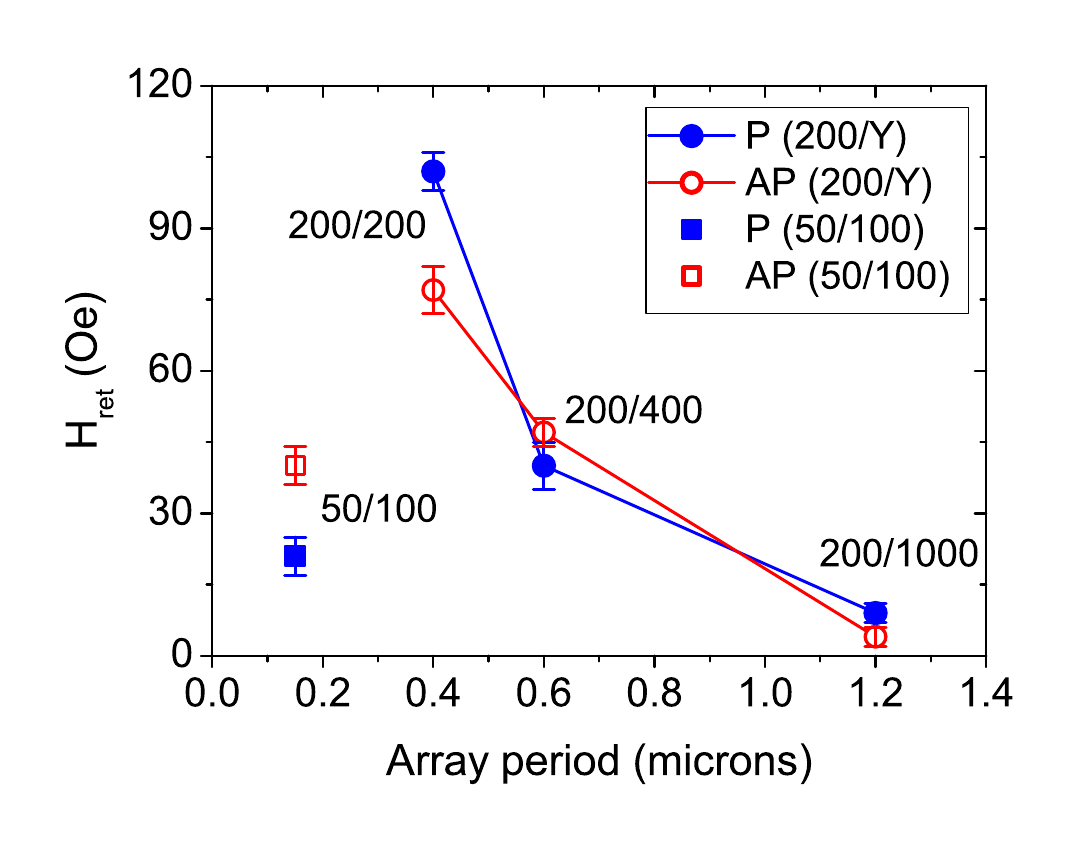}
\caption{(Color online) Retardation field values, $H_{ret}$, for the P and AP configurations as deduced from the collapsed $\ln v$ versus $(H - H_{ret})^{-1/4}$ data sets [insets of Fig.~\ref{fig_vels}]. Values are shown as a function of the array period, $L=X+Y$ [see Fig.~\ref{fig_1}(a)]. Error bars represent uncertainties in the $H_{ret}$ value which yields the best data overlap in Fig.~\ref{fig_vels}.}
\label{fig_hret}
\end{figure}

In Fig.~\ref{fig_hret} we have plotted the values of the retarding fields for the P and AP configurations, $H^P_{ret}$ and  $H^{AP}_{ret}$ versus the array period $L=X+Y$, as determined from the adjusted data in the insets of Fig.~\ref{fig_vels}. For the arrays of 200 nm wide nanoplatelets the retarding fields increase as the array period, $L$, decreases (ie.~as the density of the nanoplatelets increases). Such a dependence can be qualitatively explained by considering a point-pinning model wherein the Zeeman and elastic energies associated with a pinned, bowed wall compete with one another\cite{Yang1999,Chikazumi}, yielding a $1/L$ dependence for $H_{ret}$. Preliminary micromagnetic simulations of the depinning also yield similar $H_{ret}-L$ behavior for both nanoplatelet sizes but underestimate the $H_{ret}$ values. More refined simulations are required which would ideally include disorder within the film as well as finite temperature which is important for reproducing thermally activated creep. Although not suitable for looking at the internal structure of the Bloch wall, generalized interface models\cite{Chauve2000,Kolton2009,Lecomte2009} in which disorder and finite temperature can be included may be well suited to this simulation task.

Despite consistently rougher walls in the AP configuration for the 200/Y arrays, the average velocity of the expanding domain wall front can be faster in  the AP configuration than in the P configuration. This is the case for the 200/200 and 200/1000 arrays where  $H^P_{ret}>H^{AP}_{ret}$ ($H^{AP}_{ret}>H^{P}_{ret}$ for the 50/100  and 200/400 arrays). At this stage, we are unable to simply explain this result. This asymmetry however leads to large variations in the domain wall velocity for the two configurations within the creep regime. For example, for $H = 175$ Oe in the 200/200 array, the creep velocity in the AP configuration, $v^{AP}$, is  almost two orders of magnitude larger than for the P configuration, $v^P$, with $v^{AP}/ v^P \approx 90$.

\subsection{Dipolar biasing of magnetization reversal}\label{sec_hyst}

\begin{figure}[htbp]
\includegraphics[width=8cm,clip=yes]{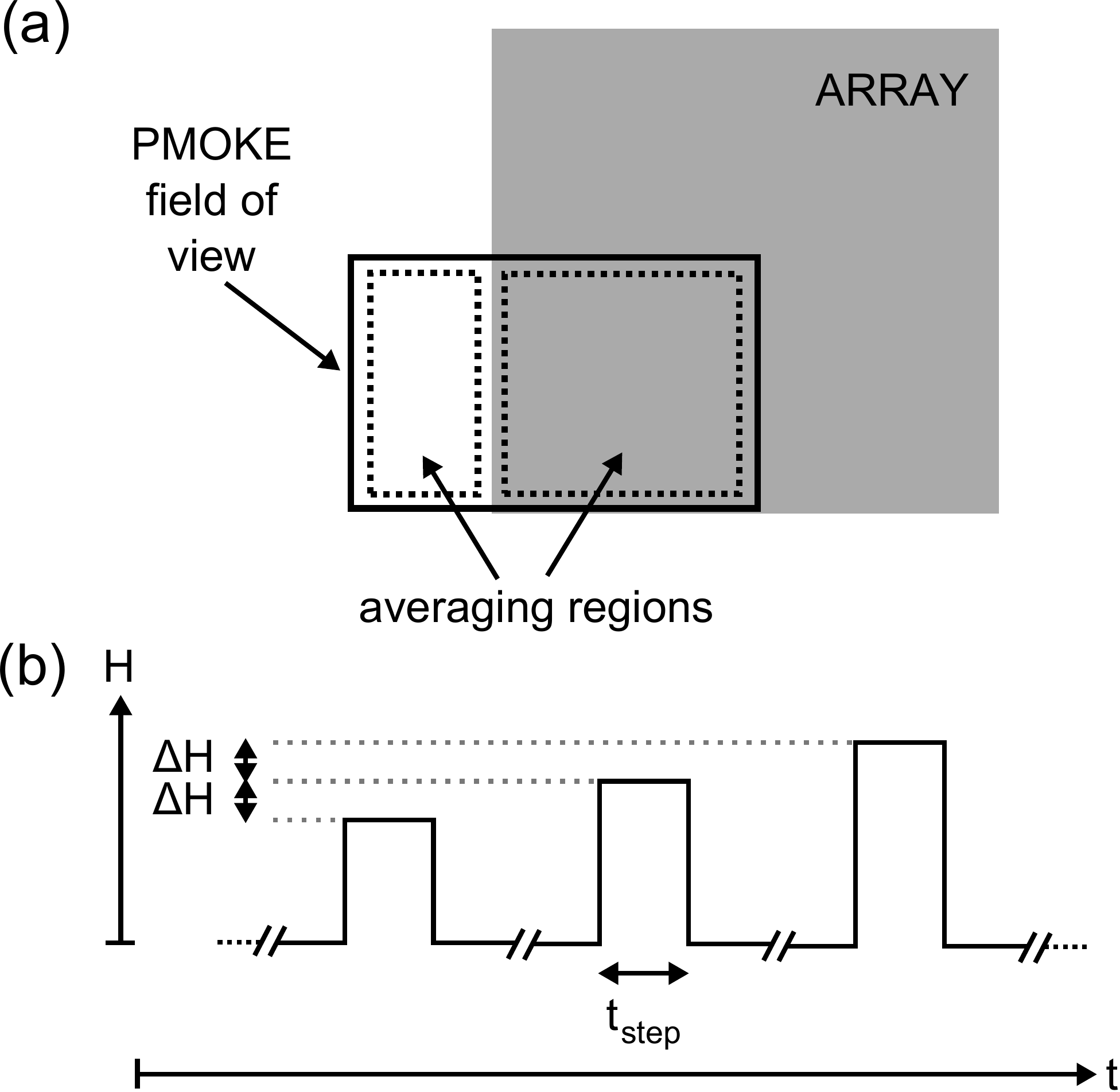}
\caption{(a)  To obtain a quasi-static remanent PMOKE hysteresis loop of the continuous layer beneath and outside the array, the system is imaged at remanence and the average PMOKE signal of the soft layer is measured in two regions: one covering a corner of the array and another covering a region directly next to the array. These signals are normalized in order to construct the $M-H$ loops shown in Fig.~9. (b) A portion of the field application sequence used to obtain the positive branch of a hysteresis loop where a field pulse is applied over a time $t_{step}$  at each step of the sequence. }
\label{fig_procedure}
\end{figure}

We now demonstrate how  asymmetric pinning ($H_{ret}^P\ne H_{ret}^{AP}$) biases magnetization reversal in the continuous layer. In this section, we examine local minor remanent hysteresis loops of the continuous layer that have been obtained  outside and underneath  [Fig.~\ref{fig_procedure}(a)] the 200/200 array for fixed $M_{array}$: \up, \down~and 0, the latter corresponding to the demagnetized, D, array configuration. The loops  were obtained by quasi-statically varying the applied field and monitoring the magnetization reversal of the soft layer using the PMOKE microscope. For each field value, a field pulse was applied over a time \tstep~[Fig.~\ref{fig_procedure}(b)] and the system imaged at remanence. The  loops were constructed by averaging and then normalizing the PMOKE signal for each image simultaneously over the two regions marked out in Fig.~\ref{fig_procedure}(a):  the  region of the film beneath the  visible part of the array  and the region of the film next to the array where there are no nanoplatelets.

\begin{figure*}[htbp]
\includegraphics[width=14cm,clip=yes]{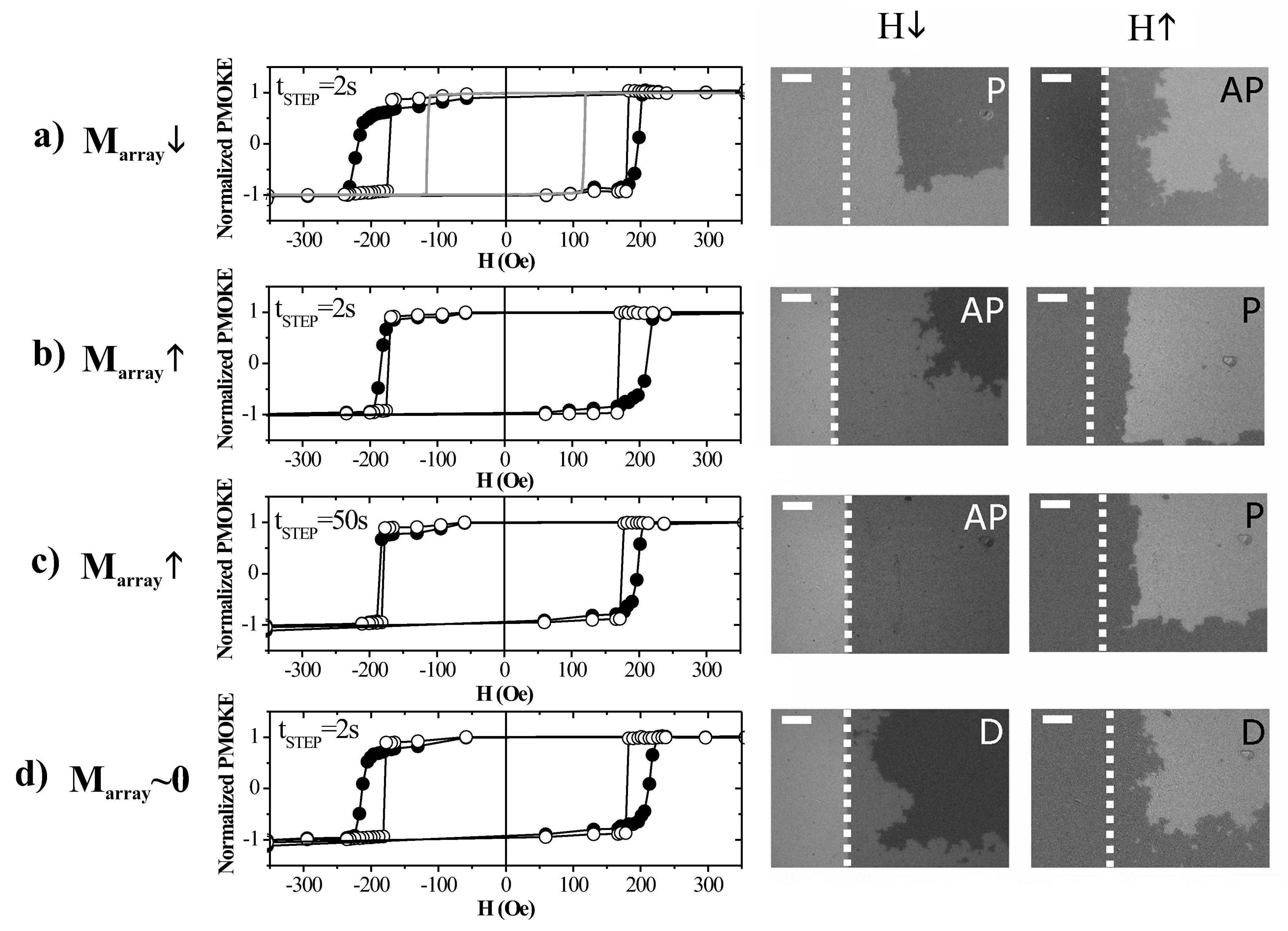}
\caption{Left hand side: Normalized, spatially averaged, remanent PMOKE hysteresis loops (black dots) of the soft layer measured in a  region located underneath the negatively (a), positively (b, c) and  demagnetized (d) 200/200 nanoplatelet array.  The equivalent loop for reversal limited only by wall pinning is shown as a gray line in (a) (see Appendix \ref{sec_creep_hyst}). The values of $t_{step}$ are indicated. Right hand side: corresponding PMOKE images for negative and positive values of the applied field close to the coercive field of the portion of the continuous layer beneath the array. The vertical dotted white line indicates the left boundary of the array [see also Fig.~8(b)]. The white scale bars are 10 $\mu$m long.
}
\label{fig_hyst}
\end{figure*}

On the left hand side of Fig.~\ref{fig_hyst} we show the remanent loops obtained beneath the array (filled circles) and next to the array (open circles)  for different values of $M_{array}$ and $t_{step}$.   On the right hand side of Fig.~\ref{fig_hyst} we show PMOKE images of domain walls moving through the array near the positive and negative coercive fields of the continuous film region located beneath the array. The dotted white line on the PMOKE images  marks the left hand boundary of the visible part of the array [the bottom left portion of  array lies to the right of the white line,  see Fig.~\ref{fig_procedure}(a)]. The fields applied during the loops were kept below the nanoplatelet switching fields to ensure a stable \marray~value. 

Loops obtained outside the array are symmetric around zero field within experimental uncertainty (related to the field step, $\Delta H$) since there are no nanoplatelets and thus no source of pinning asymmetry. Such loops  can be used to monitor the reversal of the magnetization in the visible region of the film just outside the array and allow us to determine the field at which a domain wall arrives at the array edge. This process is initiated by nucleation which occurs outside the field of view of the microscope and results in  a coercive field for this region of $175\pm 5$ Oe. 

To reproducibly prepare the wall next to the array, the same field sequence is used for each loop up until the arrival of the wall at the array edge. Note that the wall always enters the field of view and arrives at the array edge during a single field step (ie.~it does not sweep slowly across the visible region outside the array over multiple field steps). For each branch of the loop, once the domain wall has arrived at the array edge, we ramp the field quasi-statically with the magnitude of subsequent field pulses increasing or decreasing by $\Delta H= 5$ Oe  [Fig.~\ref{fig_procedure}(b)]. In this phase of the loop, each field pulse is applied for a time \tstep~which is chosen for each loop. The PMOKE microscope field of view  covers the bottom, left hand corner of the array [Fig.~\ref{fig_procedure}(a)]. Since domain walls originate at the array edges in these hysteresis measurements, the walls are typically seen propagating  both upwards and toward the right when moving through the arrays (Fig.~\ref{fig_hyst}).

As the domain wall sweeps through a magnetized array (filled circles in Figs.~\ref{fig_hyst}(a-c) with \arrayup~or \arraydown), it experiences a pinning potential which depends upon the relative direction of the field and the fixed array magnetization.  In Fig.~\ref{fig_hyst}(a), the loop is obtained under the negatively saturated array, \arraydown. For \arraydown, the domain wall is subject to P-pinning for $H<0$ (\hdown) and   AP-pinning for $H>0$ (\hup). Consistent with the wall motion being slowest in the P configuration for this array [Fig.~6(b)]  the switching field  in Fig.~\ref{fig_hyst}(a) is largest in the negative branch, resulting in the observed bias shift  towards negative field values ($H_B = -13 \pm  5$ Oe).  When switching the nanoplatelet's magnetization to \arrayup~[Fig.~\ref{fig_hyst}(b)], the P configuration occurs for $H>0$ (\hup) and as a result the loop is shifted toward positive field values. The direction of the shift is thus switchable, being determined by \marray~[compare Figs.~\ref{fig_hyst}(a,b)].

The `mean' coercivity of the soft layer below the array, $\bar{H}_C= 202 \pm  8$ Oe, 	is also increased with reference to the region outside. This is a result of the periodic pinning potential slowing down wall motion for both the P and AP configurations. Finally, we note that the domain wall roughness near coercivity is higher in the AP branch of each loop  [Fig.~\ref{fig_hyst}(a-c)] which is consistent with the images in Fig.~\ref{fig_moke}(d,e) where domain morphologies beneath the P- and AP-configured 200/200 array can be directly compared.

Lowering the effective sweep rate ($R_{eff}=\Delta H/t_{step}$) reduces the bias field. This can be seen in Fig.~\ref{fig_hyst}(c) where the use of a longer $t_{step}$ of $50$ s leads to a reduced bias field of $H_B=6\pm 5$ Oe. This rate dependence can be reproduced well by simply calculating the change in the normalized PMOKE signal due to two straight walls moving across the array from the bottom of left hand sides  using the experimentally obtained velocity-field response [Fig.~\ref{fig_vels}(c)] and a  field which is swept continuously at $R_{eff}$ from a starting value of $\pm175$ Oe. The calculation yields $H_B=5.2$ Oe at $t_{step}=50$ s and $H_B=12.2$ Oe at $t_{step}=2$ s which compare well with the measured values.

By exploiting the asymmetry of the domain wall pinning that occurs when switching the applied field direction, we can thus reproduce many effects which are typical of conventional layered ferromagnet/antiferromagnet exchange bias systems. These include a bias shift which is switchable\cite{Miltenyi1999} and  field-sweep-rate dependent\cite{Xi2001,Sahoo2007,Malinowski2007}, coercivity enhancement\cite{Leighton2000,Gokemeijer2001,Stiles2001} and asymmetry in the reversal mode itself \cite{Nikitenko2000,Kirilyuk2002,Blomqvist2005,McCord2008} (ie.~differing domain wall roughnesses on the two branches of the hysteresis loop).   In conventional exchange bias systems both rate dependence of the bias field\cite{Stamps2000,Xi2001,Sahoo2007} and coercivity enhancement\cite{Stamps2000,Stiles2001,Scholten2005}  have  been linked to processes within the antiferromagnet. However, a change in the dominant  reversal mechanism of the ferromagnet\cite{Malinowski2007} has also been shown to lead to rate dependent bias fields and  coercivity enhancement can be a consequence of inhomogeneities in the effective interfacial coupling fields\cite{Leighton2000,Stiles2001}. Here, both  bias and coercivity enhancement are  inhomogeneity-driven, arising due to field-polarity-dependent, nanoplatelet-induced, spatially non-uniform domain wall pinning potentials.  Furthermore, the weak rate dependence of the bias field in our system can be linked directly to the thermally activated domain wall creep and the way in which it is modified by the periodic pinning potential (calculation results noted in the text above). Finally we note that the observed coercivity enhancement distinguishes our results from  simple shifted loops that are commonly observed in coupled, continuous ferromagnet/non-magnet/ferromagnet (FM/NM/FM) multilayers. [\textit{Note added after publication in J. Appl. Phys.:} Gottwald \textit{et al}\cite{Gottwald2012} observed asymmetric reversal of the magnetically soft FM1 layer of a perpendicularly magnetized FM1/NM/FM2 trilayer when the magnetically harder FM2 layer was in a non-saturated state. This asymmetry could be linked to field-polarity-dependent modifications to domain nucleation and  wall propagation which were induced by stray fields generated by the non-uniform domain structure in FM2.]

When the array is demagnetized, we no longer observe a clear dipolar biasing effect however the coercivity remains enhanced with $H_C^D = 212  \pm 5$ Oe [Fig.~\ref{fig_hyst}(d)]. The individual nanoplatelets still locally pin the domain walls [ie.~Fig.~5c] which leads to loop broadening. However, when averaged over an ensemble of \up~and \down~ nanoplatelets, the pinning effects are independent of the applied field polarity.  Analogous coercivity enhancement in the absence of loop bias can be observed in compensated\cite{Gokemeijer2001} and frustrated\cite{Leighton2000} ferromagnet/antiferromagnet systems.   $H_C^D$ is close to the coercive field seen in the P configuration, $H_C^P = 215 \pm  5$ Oe [Figs.~\ref{fig_hyst}(a,b)] and this is consistent with the observation that the dynamics for the P and D configurations are  very similar [Fig.~\ref{fig_vels}(c)]. However, the domain structure in the D configuration [Fig.~\ref{fig_hyst}(d,right)] appears to be dominated by the AP-configured nanoplatelets since the domain morphology mirrors that observed for the AP-configured array  [Figs.~4(e) and 9(a-c)].

One feature of the  exchange bias phenomenon which has not  been reproduced here is training \cite{Nowak2002,Hoffmann2004,Binek2006} however analogous effects could potentially appear if  nanoplatelets were engineered such that their magnetic states could be perturbed by the stray fields generated by the domain walls\cite{Wiebel2005,Wiebel2006,Baruth2006} in the underlying layer.

\section{Conclusion}

Strong dipolar fields localized beneath and around magnetically hard ferromagnetic nanoplatelets can be used to locally impede domain wall motion in an underlying ferromagnetic layer. 
This offers an attractive way to study the effects of co-existing periodic\cite{Kolomeisky1995,Ettouhami2003} and disordered\cite{Lemerle1998} pinning potentials on domain wall motion, especially when noting the applicability of domain walls in ultrathin Pt/Co/Pt layers to the rich, fundamental problem of elastic interface dynamics in disordered media\cite{Lemerle1998,Metaxas2007,Kim2009}. 

Pinning in  these systems depends on the relative direction of the applied field and the magnetization of the nanoplatelets. For a given $M_{array}$, this leads to domain wall roughness and velocity which are field polarity dependent, allowing for a degree of pinning tunability. A further consequence however is that domain wall mediated switching in the continuous layer is  biased with this purely ferromagnetic system exhibiting many phenomena that are normally associated with conventional ferromagnet/antiferromagnet exchange bias systems. Indeed, this work demonstrates that many common features of the exchange bias phenomenon can be  reproduced experimentally simply by the addition of a spatially non-uniform local field whose effects upon reversal depend upon the polarity of the applied field.

\begin{acknowledgments}
P.J.M. acknowledges an Australian Postgraduate Award and a Marie Curie Action (MEST-CT-2004-514307). P.J.M., J.F., A.M. and R.L.S. acknowledge support from the FAST (French-Australian Science and Technology) program. R.L.N. and A.M. are grateful for financial support from the French ANR-08-NANO-P196-36 MELOIC.  P.J.Z., G.G., V.B. and B.R. acknowledge the support of NanoFab/CNRS for film patterning. This research was partly supported by the Australian Research Council's Discovery Projects funding scheme (project number DP0988962).
\end{acknowledgments}

\appendix

\section{Micromagnetic simulation}\label{micromag-appendix}

Micromagnetic simulations were carried out using OOMMF \cite{oommf} by simulating the quasi-static, field driven  motion of a domain wall through a 0.65nm thick Co film which is free of structural defects. The film lies in the $x-y$ plane (Fig.~3) and is subject to a variable applied field $(0,0,H)$ as well as the calculated stray field, $(H_{stray}^x,H_{stray}^y,H_{stray}^z)$,  of  the nanoplatelets which would surround and be contained within the lateral calculation region. The domain wall moves in the $x$-direction and periodic boundary conditions\cite{Lebecki2008} were used in the $y$ direction. The following simulation parameters were used. Cell size: $2.5 \times 2.5 \times 0.65$ nm$^3$, damping parameter: $\alpha= 0.5$, saturation magnetization: $M_S = 1.165 \times 10^6$ A/m, exchange constant: $A = 1.7\times 10^{-11}$ J/m, uniaxial out of plane anisotropy energy density: $K = 1.24 \times 10^6 $ J/m$^3$, stopping criterion: $|dm/dt|<0.1^\circ$/ns.

\section{PMOKE: Domain preparation, expansion and velocity extraction}\label{moke-appendix}

 For the PMOKE experiments, domain walls were driven under positive field, \hup, after setting \marray. Setting \marray~and preparing a positively magnetized domain entailed three steps for the P and D configurations and 2 steps for the AP configuration. In the P configuration, the continuous layer and nanoplatelets were first positively magnetized (\arrayup) under +4 kOe (P step 1). A negative field of about -1 kOe was then applied to negatively magnetize the continuous layer (P step 2). Short positive field pulses ($\sim$1 kOe over $\sim$100 ns) were then used to nucleate reversed, positively magnetized domains within the array which could be expanded under applied fields via domain wall propagation (P step 3).  For the AP configuration, the continuous layer and nanoplatelets were first negatively magnetized (\arraydown) under -4 kOe (AP step 1).  Short positive field pulses were then used to nucleate positively magnetized domains in the continuous layer which were expanded under \hup~(AP step 2 = P step 3). For the D configuration, the array was demagnetized by saturating the nanoplatelets in 4 kOe and then applying a field with opposite polarity and  a magnitude that was chosen such that half of the dots reversed. P steps 2 and 3 were then carried out. 
 
  Images of both the initially nucleated domain and the field-expanded domain within the continuous layer were obtained at remanence using the PMOKE microscope. The area swept out by the domain wall front (and thus the front's displacement) was determined by subtracting the two images. The velocity was then determined by dividing the wall displacement by the pulse length. For short pulses, where the rise time  was comparable to the total pulse length, a multiple pulse method was used which removed effects from the transient parts of the pulse\cite{Metaxas2007}.

\section{MFM: Imaging and domain preparation and  expansion}\label{mfm-appendix}

MFM imaging was carried out with a Veeco/Digital Instruments Dimension 3100 using a CoCr coated tip with a resonant frequency on the order of 70 kHz. The CoCr sputtering time for the deposition of the CoCr was chosen to provide a tip which would give good contrast while minimizing interactions with the sample and associated tip-induced domain wall displacement. Interleaved scans at $\sim$1 line/s were used with a tip height of 30 nm.

Domain preparation for MFM imaging was performed using similar procedures to those described for PMOKE microscopy except that (i) for the P and AP measurements, the array was in the \arrayup~state and \hup~and \hdown~fields were respectively used to obtain the P and AP configurations, (ii) domains were nucleated outside the arrays and then injected inside the arrays using fields on the order of 100 Oe and (iii) to prepare the D state,  the sample was submitted to a slow oscillating field with decreasing amplitude (from $\pm$4 kOe to 0 Oe).

\section{Wall-limited coercivity}\label{sec_creep_hyst}

The gray loop in Fig.~\ref{fig_hyst}(a,left) is that obtained by preparing a wall just outside the field of view of the PMOKE microscope  and  then stepping the field from $H=0$ ($t_{step} = 2$ s and  $\Delta H =  2$ Oe) while measuring a spatially averaged remanent hysteresis loop in a region without nanoplatelets. Despite nucleation not being the defining factor for the onset of reversal, a finite coercivity is still observed due to domain wall pinning by intrinsic structural inhomogeneities in the continuous layer. This wall-limited coercivity is $117 \pm 2$ Oe. Although noticeably smaller than the nucleation-limited coercive field of $175\pm 5$ Oe obtained in the region just next to the arrays, caution must be taken when directly comparing the two values since different field application sequences have been used.

\end{document}